\documentclass[prl, twocolumn, superscriptaddress]{revtex4}

\usepackage{graphicx}
\usepackage{amsmath}
\usepackage{amsfonts}

\renewcommand{\vec}{\boldsymbol}

\usepackage{color}

\begin{document}

\title{Real-time dynamics of the Chiral Magnetic Effect}

\author{Kenji Fukushima}
\affiliation{Yukawa Institute for Theoretical Physics,
 Kyoto University, Kyoto 606-8502, Japan}

\author{Dmitri E. Kharzeev}
\affiliation{Department of Physics,
Brookhaven National Laboratory, Upton NY 11973, USA}

\author{Harmen J. Warringa}
\affiliation{Institut f\"ur Theoretische Physik,
 Goethe-Universit\"at Frankfurt,
 Max-von-Laue-Stra\ss e~1, 60438 Frankfurt am Main, Germany}


\begin{abstract}
  In quantum chromodynamics, a gauge field configuration with nonzero
  topological charge generates a difference between the number of
  left- and right-handed quarks. When a (electromagnetic) magnetic
  field is added to this configuration, an electromagnetic current is
  induced along the magnetic field; this is called the chiral magnetic
  effect. We compute this current in the presence of a color flux tube
  possessing topological charge, with a magnetic field applied
  perpendicular to it. We argue that this situation is realized at the
  early stage of relativistic heavy-ion collisions.
\end{abstract}

\maketitle
{\it Introduction.} 
The theory of the strong interactions, quantum chromodynamics (QCD),
is an $\mathrm{SU(3)}$ Yang-Mills theory coupled to fermions
(quarks). An intriguing aspect of $\mathrm{SU}(N)$ Yang-Mills theories
is their relation to topology. This reveals itself in the existence of
gauge field configurations carrying topological charge $Q$
\cite{BPST}. This charge is quantized as an integer if these
configurations interpolate between two of the infinite number of
degenerate vacua of the $\mathrm{SU}(N)$ Yang-Mills theory
\cite{CDG}. Expressed in terms of the field strength tensor $G^{\mu
  \nu}_a$ the topological charge reads $Q = \frac{g^2}{32\pi^2} \int
\mathrm{d}^4 x \, G^{\mu \nu}_a \tilde G_{\mu \nu}^a$; here $g$
denotes the coupling constant and the dual field strength tensor
equals $\tilde G^{\mu \nu a} = \tfrac{1}{2} \epsilon^{\mu \nu \rho
  \sigma} G^a_{\rho \sigma}$.

By interacting with fermions the $Q\neq 0$ fields induce parity
$(\mathcal{P})$ and charge-parity ($\mathcal{CP}$) odd effects
\cite{H}. This can be seen by the following exact equation (valid for
each quark flavor $\psi$ separately) which is a result of the
$\mathrm{U}(1)$ axial anomaly \cite{S51,ABJ}: $\partial_\mu j_5^\mu =
2 m \langle \bar \psi \gamma^5 \psi \rangle_A - \frac{g^2}{16\pi^2} \,
G^{\mu \nu}_a \tilde G_{\mu \nu}^a,$ where $m$ is the quark mass, and
$j_5^\mu = \langle \bar \psi \gamma^\mu \gamma^5 \psi \rangle_A$
denotes the axial current density in the background of a gauge field
configuration $A_\mu^a$. Let us define the chirality { density} $n_5 =
j_5^0$ and the chirality $N_5 = \int \mathrm{d}^3 x\,
n_5$. Integrating the anomaly equation over space and time gives for
massless quarks $\Delta N_5 = -2 Q$, where $\Delta N_5$ denotes the
change in chirality over time. For massless quarks, the chirality
$N_5$ is equal to the difference between the number of particles plus
antiparticles with right-handed and left-handed helicity. Again for
$m=0$ right-handed helicity implies that spin and momentum are
parallel whereas they are antiparallel for left-handed helicity.

The $Q \neq 0$ gauge fields are included in the path integral, and as
a result they contribute to the amplitudes of physical processes.
Experimental evidence for these configurations is however indirect.
The clearest confirmation follows from the large mass of $\eta'$
pseudoscalar meson compared to the $\pi$, $K$, and $\eta$ mesons
\cite{H}. In this Letter we will discuss an alternative way in which
topological configurations of gauge fields in QCD, i.e.\ gluon fields
could be studied in heavy-ion collisions.

Using high-energy heavy-ion collisions at the Relativistic Heavy Ion
Collider (RHIC) and the Large Hadron Collider (LHC) one can
investigate the behavior of QCD at high energy densities. Very strong
color electric and color magnetic fields are produced during these
collisions, whose strength is characterized by the gluon saturation
momentum $Q_s$. In addition, extremely strong (electromagnetic)
magnetic fields are present in non-central collisions, albeit for a
very short time. In gold-gold collisions at RHIC energies the
magnitude of this magnetic field at the typical time scale $\sim
Q_s^{-1}\approx$ $0.2\;\mathrm{fm}/\mathrm{c}$ after the collision is
of the order of $\sim 10^4 \;\mathrm{MeV}^2$ which corresponds to
$\sim 10^{18}\;\mathrm{G}$ \cite{KMW, SIT}. Such extremely strong
magnetic fields are able to polarize to some degree the bulk of the
produced quarks which have typical momenta of a few hundred MeV.
More specifically, quarks with positive (negative) charge have a
tendency to align their spins parallel (anti-parallel) to the magnetic
field. As a result, assuming the produced quarks can be treated as
massless, a positively (negatively) charged quark with right-handed
helicity will have its momentum parallel (antiparallel) to the
magnetic field. For quarks with left-handed helicity this is exactly
opposite. Hence a quark and anti-quark both having the same helicity
will move in opposite directions {with respect to} the magnetic
field. This implies that an electromagnetic current is generated along
the magnetic field if there is an imbalance in the helicity, i.e. a
nonzero chirality. Because gauge fields with $Q \neq 0$ generate
chirality, they will therefore induce an electromagnetic current along
a magnetic field.  This mechanism which signals $\mathcal{P}$- and
$\mathcal{CP}$-odd interactions has been named the chiral magnetic
effect \cite{KMW, KKZ}. In an extremely strong magnetic field $\vec
B$, so strong that all quarks are fully polarized, it follows from the
arguments presented above that for each quark flavor separately the
induced current equals $\vec J = \vert q \vert N_5 \vec B / \vert \vec
B \vert = - 2 \vert q \vert Q \vec B / \vert \vec B \vert$, where $q$
is the charge of the quark.

The magnetic field in heavy ion collisions is pointing in a direction
perpendicular to the reaction plane; this is the plane in which the
impact parameter and the beam axis lie. As a result of the chiral
magnetic effect the charge asymmetry between the two sides of the
reaction plane will be generated. The sign of this asymmetry will
fluctuate from collision to collision since (assuming the so-called
$\theta$ angle vanishes and there is no global violation of parity)
the probability of generating either positive $Q$ or negative $Q$ is
equal. Using the observable proposed in \cite{V04} the STAR
collaboration has analyzed charge correlations \cite{star}.  The
results are qualitatively in agreement with the predictions of the
chiral magnetic effect; the search for alternative explanations and
additional manifestations of local parity violation is underway
\cite{WBKL}.

\begin{figure}
 \includegraphics[width=0.5\columnwidth]{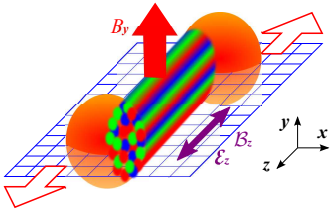}
 \caption{Schematics of the collision geometry and fields.}
 \label{fig:glasma}
\end{figure}

Several quantitative theoretical studies of the chiral magnetic effect
have appeared in the literature \cite{FKW, adscft, lattice,
  inst}. Most of the analytic studies are based on introducing a
chiral asymmetry by hand, after which the equilibrium response to a
magnetic field is studied \cite{FKW, adscft} (see also \cite{add}). In
this Letter we will for the first time investigate a situation in
which the chirality is generated dynamically in real-time in the
presence of a magnetic field.  For this we will take the simplest
Yang-Mills gauge field configuration carrying topological charge,
that is one which describes a color flux tube having constant Abelian
field strength, i.e.\ $G^{\mu \nu}_a = \mathcal{G}^{\mu \nu} n^a$ with
$n_a n_a = 1$ and $\mathcal{G}^{\mu \nu}$ constant and homogeneous.
Furthermore, we will take only the $z$-components of the color
electric ($\mathcal{E}_z = \mathcal{G}_{0 z}$) and color magnetic
($\mathcal{B}_z = -\tfrac{1}{2}\epsilon_{z i j} \mathcal{G}^{i j} $)
field nonzero. Perpendicular to this field configuration we will apply
an electromagnetic field $B_y$ pointing in the $y$ direction (see
Fig.~\ref{fig:glasma}). Note that hereafter we write $\mathcal{B}$ to
denote a \textit{color} magnetic field and $B$ for an
\textit{electromagnetic} one.  Such color flux tubes, which carry
topological charge and are homogeneous over a spatial scale $\sim
Q_s^{-1}$, naturally arise in the glasma \cite{glasma}, the dense
gluonic state just after the collision, where
$\mathcal{E}_z\sim\mathcal{B}_z\sim Q_s^2/g$. The induced current
itself can generate electromagnetic and color fields, which can alter
the dynamics. We will ignore this back-reaction, which can be
justified as long as the induced current is small compared to the
currents that create the external color and magnetic fields.
Furthermore we will also ignore the production of gluons in the color
flux-tube.

{\it Calculation. } Using a color rotation we can choose only the
third component of $n^a$ nonvanishing. Since the generator $t^3 =
\mathrm{diag}(\tfrac{1}{2}, -\tfrac{1}{2}, 0)$ of the
$\mathrm{SU(3)}$ Lie algebra is diagonal, the different color
components decouple. As a result for each quark flavor separately the
problem is equivalent to a quantum electrodynamics (QED) calculation,
in which the magnetic field $\vec B = (0, B_y, B_z)$ with $q B_z
=\pm\tfrac{1}{2}g\mathcal{B}_z$ and the electric field $\vec E = (0,
0, E_z)$ with $q E_z = \pm \tfrac{1}{2} g \mathcal{E}_z $.  Here $\pm$
labels the different color components, and $q$ denotes the electric
charge of a particular quark.  We will define $K$ to be the coordinate
frame in which the electromagnetic field has this form.

\begin{figure}[t]
\includegraphics{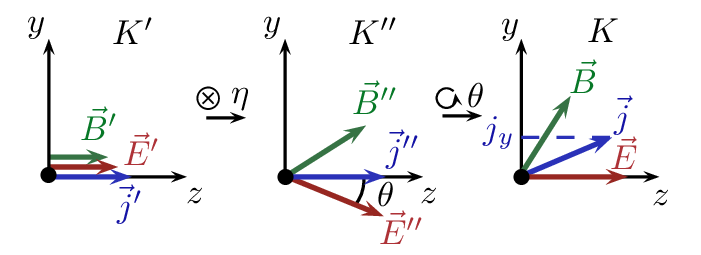}
\caption{ Lorentz transformation from a frame $K'$ in which the
    electric field ($\vec E$), magnetic field ($\vec B$), and the current
    density ($\vec j$) are parallel to each other, to a frame $K$ in
    which $\vec B$ and $\vec j$ have a component perpendicular to
    $\vec E$. 
\label{fig:lorentztransformation}
}
\end{figure}

We hence need to compute the induced electromagnetic current density
$j^\mu = q \langle \bar \psi \gamma^\mu \psi \rangle$ in $K$. To do
this we will start in a different coordinate system $K'$ in which
$\vec E = (0, 0, E_z')$ and $\vec B = (0,0,B_z')$. In this frame it is
rather straightforward to do calculations. Then by applying a Lorentz
transformation we can obtain the results in $K$ as is illustrated in
Fig.~\ref{fig:lorentztransformation}.  We will switch on the electric
field in $K'$ uniformly at a time $t'_i$ in the distant past,
i.e. $E'_z(t') = E_z' \theta(t'-t'_i)$. In this way the situation in
$K'$ is completely homogeneous.

In $K'$ particle-antiparticle pairs are produced by the Schwinger
process \cite{S51}.  The rate per unit volume of this process equals
\cite{N69}, (see also~\cite{Dunne} and \cite{CM08})
\begin{equation}
 \Gamma = \frac{q^2 E_z' B_z'}{4\pi^2} 
  \coth \left(\frac{B_z'}{E_z'} \pi \right) 
 \exp \left( - \frac{m^2 \pi}{\vert q E_z' \vert} 
\right). 
\label{eq:gamma}
\end{equation}
The production of pairs in $K'$ gives rise to an homogeneous
electromagnetic current density $j'^\mu$. Because of symmetry reasons
the only nonvanishing component of this current lies in the
$z$-direction.  Furthermore, each time a pair is created the current
will grow. Eventually when both components of the pair are accelerated
by the electric field to (nearly) the speed of light, the net effect
of the creation of one single pair will be that the total current has
increased by two units of $q$. Therefore, sufficiently long after the
switch-on, the change in current density in the $z$-direction becomes
$2q$ times the rate per unit volume of pair-production, to be precise
$\partial_{t'} \vec{j}' = 2 q \Gamma \mathrm{sgn}( q E_z') \vec
e_z$. This equation has been verified explicitly numerically in
\cite{T08}.  We have also found it to be correct analytically, even
for $m\neq 0$ \cite{F10}.

Before we compute the induced currents in $K$ let us point out that
the rate $\Gamma$ is consistent with the anomaly equation.  In the
limit of a very large magnetic field ($B_z' \gg E_z'$) all produced
pairs will reside in the lowest Landau level causing maximal chiral
asymmetry. Since each pair then produces two units of $N_5$, the pair
production rate should then be equal to half the chirality rate.
Taking the limit $B'_z \gg E'_z$ in
Eq.~(\ref{eq:gamma}) gives
\begin{equation}
\Gamma\, \mathrm{sgn}(E_z' B_z')  \approx \frac{q^2}{4\pi^2} E_z' B_z' 
\exp \left( - \frac{m^2 \pi}{\vert q E_z' \vert}  \right)
=
\tfrac{1}{2} \partial_{t'} n'_5,
\label{eq:anomaly2}
\end{equation}
which is indeed in agreement with the anomaly equation (see
Introduction) in the limit of $m=0$, since the chiral current
  $\vec j_5$ vanishes because of homogeneity. It turns out that
Eq.~(\ref{eq:anomaly2}) also exactly gives the chirality rate for
nonzero $m$ and any $E'_z$ and $B_z'$ \cite{F10}.

As is indicated in Fig.~\ref{fig:lorentztransformation} we can go from
frame $K'$ to $K''$ by applying a boost with rapidity $\eta$ in the
$x$-direction. In the new coordinate system $K''$ obtained by this
boost, the electric and magnetic field respectively read $ \vec E'' =
- B'_z \sinh \eta \, \vec e_y + E'_z \cosh \eta \, \vec e_z$ and $\vec
B'' = E'_z \sinh \eta \, \vec e_y + B'_z \cosh \eta \, \vec e_z$.
Since $j'^\mu$ points in the $z$-direction, the direction of $j'^\mu$
will not change after the boost in the $x$-direction. However because
the boost implies that $t' = t''\cosh\eta + x'' \sinh\eta$, the
current density rate is modified to $\partial_{t''} \vec{j}'' = 2 q
\Gamma \mathrm{sgn}(q E_z') \cosh\eta\, \vec e_z$.  The current
density has now also obtained a gradient in the $x$-direction
($\partial_{x''} \vec{j}'' \neq 0$). This and other inhomogeneities in
$K''$ arise because the uniform switch-on of $\vec E'$ at $t'_i$
implies an inhomogeneous switch-on of part of $\vec E''$ and $\vec
B''$ at $t''=t'_i/\cosh\eta -x''\tanh\eta$.

To arrive in frame $K$ we have to apply a rotation with angle $\theta$
around the $x$-axis such that the electric field points in the
$z$-direction. The angle $\theta$ follows from
Fig.~\ref{fig:lorentztransformation} and\ satisfies $\sin \theta =
-E_y''/E_z = B_z' \sinh\eta / E_z$ and $\cos \theta = E_z''/E_z = E_z'
\cosh\eta / E_z$.  The current density rate becomes after the rotation
\begin{equation}
\partial_t \vec j =
 q \Gamma
\left(\sinh(2\eta) \, \frac{B_z'}{E_z} \vec e_y
+ \cosh^2\eta\,
\frac{2 E_z'}{E_z} \, \vec e_z \right) \mathrm{sgn}(q E_z').
\end{equation}
We can eliminate $\eta$ by expressing the above in terms of the fields
in $K$.  The magnetic field is $B_y = E_z' \sinh\eta\, \cos \theta +
B_z' \cosh\eta\, \sin \theta$, implying that $\sinh(2\eta) = 2 B_y E_z
/ (E_z'^2 + B_z'^2)$.  Because both $\mathcal{F} = \tfrac{1}{4} F_{\mu
  \nu} F^{\mu \nu} = \tfrac{1}{2} (B_y^2 + B_z^2 -
E_z^2)=\tfrac{1}{2}(B_z'^2-E_z'^2)$ and $ \mathcal{H} = - \tfrac{1}{4}
F_{\mu \nu} \tilde F^{\mu \nu} = E_z B_z = E_z' B_z'$ are Lorentz
invariant, one finds $a \equiv \vert E_z' \vert = (
\sqrt{\mathcal{F}^2 + \mathcal{H}^2} - \mathcal{F} )^{1/2}$, and $b
\equiv \vert B_z' \vert = ( \sqrt{\mathcal{F}^2 + \mathcal{H}^2} +
\mathcal{F})^{1/2}$.

Now we can put all our results together. After summing over colors the
$z$-component of the current vanishes ($\partial_t j_z=0$), implying
that the only remaining component lies in the $y$-direction. Using
that $q \, \mathrm{sgn}(q E_z') B_z' = \vert q \vert
\mathrm{sgn}(\mathcal{E}_z \mathcal{B}_z ) b$ we obtain
after summing over colors,
\begin{equation}
\partial_t j_y
= \frac{q^2 \vert q \vert B_y}{\pi^2} 
\frac{a b^2 \mathrm{sgn}( \mathcal{E}_z \mathcal{B}_z )}{a^2 + b^2}
\coth \left(\frac{\pi b}{a} \right) 
\exp\biggl( - \frac{m^2 \pi}{\vert q a \vert}\biggr)
\label{eq:djydt}
\end{equation} 
where $a$ and $b$ have dependence on $q E_z = \pm \tfrac{1}{2} g
\mathcal{E}_z$ and $q B_z = \pm \tfrac{1}{2}g \mathcal{B}_z$.  The
rate of chirality production in $K$ becomes $\partial_t n_5 =
\cosh^2\eta\, \partial_{t'} n'_5$.  Inserting Eq.~(\ref{eq:anomaly2})
yields for the rate of current over chirality density generation
\begin{equation}
\frac{1}{\vert q \vert }
\frac{\partial_t j_y}
{\partial_t n_5}
=
\frac{2 q^2 B_y b \coth \left(\pi b / a \right)}
{q^2(a^2 + b^2 + B_y^2) + 
\tfrac{1}{4} g^2 (\mathcal{E}_z^2 + \mathcal{B}_z^2)
}.
\label{eq:djydn5}
\end{equation}

\begin{figure}[t]
\includegraphics{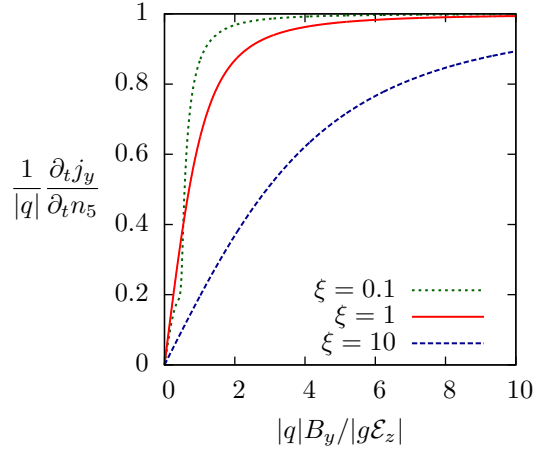}
\caption{ Rate of current ($j_y$) over chirality density ($n_5$) generation in a
    color flux tube, as a function of the perpendicular magnetic field
    $B_y$. The ratio $\xi = \vert \mathcal{B}_z /
  \mathcal{E}_z \vert$. The curves are valid for any value of the quark mass.
  \label{fig:djydn5}}
\end{figure}

{\it Discussion. }  Equation~(\ref{eq:djydt}) clearly shows that an
external magnetic field induces a current perpendicular to the color
flux tube. To summarize our findings we display in
Fig.~\ref{fig:djydn5} for three different values of $\xi = \vert
\mathcal{B}_z / \mathcal{E}_z \vert$ the rate of generation of this
current normalized to Eq.~(\ref{eq:djydn5}), the rate of chirality
production.  We will now analyze our results and show that $\partial_t
j_y$ indeed behaves as the chiral magnetic effect predicts.

First of all let us take either $\mathcal{E}_z = 0$ or $\mathcal{B}_z
= 0$, which implies that no chirality is generated. If
$\mathcal{E}_z = 0$ then $a=0$, for $\mathcal{B}_z= 0$ either $a=0$ or
$b=0$. In all these cases $\partial_t j_y$ indeed vanishes as follows
from Eq.~(\ref{eq:djydt}). This is obvious when $a=0$ since in
that case no particles are produced as follows from
Eq.~(\ref{eq:gamma}). Also as expected $\partial_t j_y$ vanishes if
there is no perpendicular magnetic field which can be seen from
Fig.~\ref{fig:djydn5} as well.

Secondly, in the limit of $q B_y \gg g \mathcal{E}_z, g
\mathcal{B}_z$, we have $b \simeq \vert B_y \vert$ so that from
Eq.~(\ref{eq:djydn5}) it follows that $\partial_t j_y = \vert q \vert
\mathrm{sgn}(B_y) \partial_t n_5$. This indicates that for large
magnetic fields the current rate is indeed exactly given by the
chirality rate in agreement with the prediction outlined in the
introduction. Therefore the curves in Fig.~\ref{fig:djydn5} approach
unity for when both $q B_y/g\mathcal{E}_z$ and $qB_y/ (g\mathcal{E}_z
\xi)$ are large.

A finite mass reduces the chirality and indeed also $\partial_t j_y$
as can be seen from Eq.~(\ref{eq:djydt}). In fact
Eq.~(\ref{eq:djydn5}) shows for any value of the mass the current is
proportional to the chirality. Hence the curves displayed in
Fig.~\ref{fig:djydn5} are independent of mass. Moreover let us point
out that the direction of the current is independent of the sign of
the quark charge, but does depend on the direction of the magnetic
field and the sign of the chirality, i.e.~$\mathrm{sgn}(\mathcal{E}_z
\mathcal{B}_z$). For $q B_y$ small compared to both $g \mathcal{E}_z$ and $g
\mathcal{B}_z$, we have $a \simeq \vert \tfrac{g}{2q} \mathcal{E}_z
\vert$ and $b \simeq \vert \tfrac{g}{2q} \mathcal{B}_z \vert$ so that
\begin{equation}
  \partial_t j_y
  \simeq \frac{q^2 B_y}{2 \pi^2} 
  \frac{g  \mathcal{E}_z \mathcal{B}_z^2 }
  {\mathcal{B}_z^2 + \mathcal{E}_z^2}
  \coth \left(\frac{\mathcal{B}_z} {\mathcal{E}_z }
    \pi \right) 
  \exp \left(- \frac{2 m^2 \pi}{\vert g \mathcal{E}_z \vert} \right).
\label{eq:jysmallby}
\end{equation}
The linear dependence on $B_y$ for small $B_y$ is clearly visible in
Fig.~\ref{fig:djydn5}.  The small kink at $q B_y / g \mathcal{E}_z
\simeq 1/2$ and $\xi = 0.1$ is due to the fact that $a$ and $b$ vary
rapidly around $\mathcal{F}=0$ when $ \vert \mathcal{H} \vert$ is
small compared to $\vert \tfrac{g}{2q} \mathcal{B}_z \vert^2$,
which is equivalent to $\xi \ll 1$.

The generation of a current by the transformation from frame $K'$ to
$K$ is a very general result of Lorentz invariance, and is equivalent
to the Lorentz force in frame $K'$. Therefore any charged colored
particle that is present in the color flux-tube plus magnetic field
background will experience a force in the $y$-direction if
$\mathcal{E}_z \mathcal{B}_z \neq 0$. To illustrate this we can
consider the whole calculation for fictional colored and electrically
charged scalar particles. In that case there is no anomaly so that no
chirality is generated. The results for scalar fermions can be
obtained by replacing $\coth( \pi b / a)$ by $1/[2\sinh(\pi b / a)]$
in Eq.~(\ref{eq:gamma}) \cite{Dunne} and subsequently in all other
equations. The ratio between the scalar and fermion current density
rate becomes simply $1/[2 \cosh(\pi b/a)]$, which is approximately
$1/[2 \cosh(\pi \xi)]$ for $q B_y \lesssim g \mathcal{E}_z/2$ and
$1/[2 \cosh(\pi (2 qB_y)^2 / (g^2\mathcal{E}_z \mathcal{B}_z)) ]$ for
$q B_y \gtrsim g \mathcal{E}_z/2$.  Clearly scalar particles behave
completely different from the predictions of the chiral magnetic
effect, moreover the scalar contribution to $j_y$ is always smaller
than that of fermions and even exponentially suppressed for $q B_y
\gtrsim g \mathcal{E}_z / 2$.

Let us finally stress that our quantitative results are strictly
speaking only valid for the rather special inhomogeneous switch-on of
the fields in the color-flux tube.  Nevertheless, as is the case at
later times in heavy-ion collisions, if $B_y$ is small compared to the
color fields, the effects of the inhomogeneous switch-on are
marginal. Therefore it is very likely that the result for small $B_y$,
Eq.~(\ref{eq:jysmallby}), is also correct for a homogeneous switch-on.
To further address this issue one can start from an inhomogeneous
switch-on in $K'$ that becomes homogeneous in $K''$.  However, this
situation is more complicated, at present we are unfortunately unable
to solve it exactly.

To conclude, we have shown by a dynamical calculation that if
topological charge is present in a magnetic field, an electromagnetic
current will be generated along this magnetic field. This very natural
mechanism is called the chiral magnetic effect and signals
$\mathcal{P}$- and $\mathcal{CP}$-odd interactions. As such it could
be an explanation for the charge correlations in heavy-ion collisions
observed by the STAR collaboration \cite{star}.  For a broader range
of physics applications the intermediate QED results might be of use
for testing the axial anomaly with strong field lasers.

{\it Acknowledgments. } We are grateful to Antti Gynther, Larry
McLerran, Anton Rebhan and Andreas Schmitt for discussions.  The work
of H.J.W.\ was supported by the Alexander von Humboldt Foundation.
K.F.\ was supported by the Japanese MEXT grant No.\ 20740134 and also
in part by the Yukawa International Program for Quark Hadron Sciences.
This manuscript has been authored under Contract
No.~\#DE-AC02-98CH10886 with the U.S.\ Department of Energy.

\appendix


\begin{thebibliography}{99}


\bibitem{BPST}
  A.~A.~Belavin, A.~M.~Polyakov, A.~S.~Shvarts and Yu.~S.~Tyupkin,
  Phys.\ Lett.\  B {\bf 59}, 85 (1975).

\bibitem{CDG}
  R.~Jackiw and C.~Rebbi,
  Phys.\ Rev.\ Lett.\  {\bf 37}, 172 (1976);
  C.~G.~Callan, R.~F.~Dashen and D.~J.~Gross,
  Phys.\ Lett.\  B {\bf 63}, 334 (1976).

\bibitem{H}
  G.~'t Hooft,
  Phys.\ Rev.\ Lett.\  {\bf 37}, 8 (1976);
  G.~'t Hooft,
  Phys.\ Rev.\  D {\bf 14}, 3432 (1976).



\bibitem{S51}
  J.~S.~Schwinger,
  Phys.\ Rev.\  {\bf 82}, 664 (1951).

\bibitem{ABJ}
 S.~L.~Adler,
 Phys.\ Rev.\  {\bf 177}, 2426 (1969); 
J.~S.~Bell and and R.~Jackiw,
Nuovo\ Cim.\  {\bf A60},~47~(1969).

\bibitem{KMW}
  D.~E.~Kharzeev, L.~D.~McLerran and H.~J.~Warringa,
  Nucl.\ Phys.\  A {\bf 803}, 227 (2008).

\bibitem{SIT}
  V.~Skokov, A.~Y.~Illarionov and V.~Toneev,
  Int.\ J.\ Mod.\ Phys.\  A {\bf 24}, 5925 (2009).

\bibitem{KKZ}
  D.~Kharzeev,
  Phys.\ Lett.\ B {\bf 633}, 260 (2006); 
  D.~Kharzeev and A.~Zhitnitsky,
  Nucl.\ Phys.\  A {\bf 797}, 67 (2007);




\bibitem{FKW}
  K.~Fukushima, D.~E.~Kharzeev and H.~J.~Warringa,
  Phys.\ Rev.\  D {\bf 78}, 074033 (2008);
  D.~E.~Kharzeev and H.~J.~Warringa,
  Phys.\ Rev.\  D {\bf 80}, 034028 (2009);
  K.~Fukushima, D.~E.~Kharzeev and H.~J.~Warringa,
  arXiv:0912.2961 [hep-ph].


\bibitem{lattice}
  P.~V.~Buividovich, M.~N.~Chernodub, E.~V.~Luschevskaya and M.~I.~Polikarpov,
  Phys.\ Rev.\  D {\bf 80}, 054503 (2009);
  M.~Abramczyk, T.~Blum, G.~Petropoulos and R.~Zhou,
  arXiv:0911.1348 [hep-lat].

\bibitem{adscft}
  H.~U.~Yee,
 JHEP {\bf 0911}, 085 (2009);
  A.~Rebhan, A.~Schmitt and S.~A.~Stricker,
  arXiv:0909.4782 [hep-th].



\bibitem{inst}  S.~i.~Nam,
  Phys.\ Rev.\  D {\bf 80}, 114025 (2009).


\bibitem{add}
  H.~B.~Nielsen and M.~Ninomiya,
  Phys.\ Lett.\  B {\bf 130}, 389 (1983);
  A.~Y.~Alekseev, V.~V.~Cheianov and J.~Frohlich,
   Phys.\ Rev.\ Lett.\ {\bf 81}, 3503 (1998);
  M.~A.~Metlitski and A.~R.~Zhitnitsky,
  Phys.\ Rev.\  D {\bf 72}, 045011 (2005);
  G.~M.~Newman and D.~T.~Son,
  Phys.\ Rev.\  D {\bf 73}, 045006 (2006);
  J.~Charbonneau and A.~Zhitnitsky,
  arXiv:0903.4450 [astro-ph.HE];
  E.~V.~Gorbar, V.~A.~Miransky and I.~A.~Shovkovy,
  Phys.\ Rev.\  C {\bf 80}, 032801 (2009).




\bibitem{V04}
  S.~A.~Voloshin,
  Phys.\ Rev.\  C {\bf 70}, 057901 (2004).

\bibitem{star}
  B.~I.~Abelev {\it et al.}  [STAR Collaboration],
  Phys.\ Rev.\ Lett.\ {\bf 103}, 251601 (2009);
  B.~I.~Abelev {\it et al.}  [STAR Collaboration],
  arXiv:0909.1717 [nucl-ex].

\bibitem{WBKL}
  F.~Wang,
  arXiv:0911.1482 [nucl-ex];
  R.~Millo and E.~Shuryak,
  arXiv:0912.4894 [hep-ph];
A.~Bzdak, V.~Koch and J.~Liao,
arXiv:0912.5050 [nucl-th].

\bibitem{glasma}
D.~Kharzeev, A.~Krasnitz and R.~Venugopalan,
Phys.\ Lett.\ B {\bf 545}, 298 (2002);
  T.~Lappi and L.~McLerran,
  Nucl.\ Phys.\  A {\bf 772}, 200 (2006);
D.~Kharzeev, E.~Levin and K.~Tuchin,
  Phys.\ Rev.\  C {\bf 75}, 044903 (2007).


\bibitem{N69}
  A.~I.~Nikishov,
  Zh.\ Eksp.\ Teor.\ Fiz.\  {\bf 57}, 1210 (1969);
  F.~V.~Bunkin and I.~I.~Tugov,
  Sov. Phys. Dokl., {\bf 14}, 678 (1970).

\bibitem{Dunne}
  G.~V.~Dunne,
  arXiv:hep-th/0406216.




\bibitem{CM08}
  T.~D.~Cohen and D.~A.~McGady,
  Phys.\ Rev.\  D {\bf 78}, 036008 (2008).

\bibitem{T08}
  N.~Tanji,
  Annals Phys.\  {\bf 324}, 1691 (2009).

\bibitem{F10}
  H.J.~Warringa, in preparation.




\end{thebibliography}
\end{document}